\documentclass[twocolumn, amssymb, superscriptaddress, aps]{revtex4-2}
\usepackage{graphicx}
\usepackage[T1]{fontenc}
\usepackage{lmodern}
\usepackage{color}
\usepackage{natbib}
\usepackage{amsmath}
\usepackage{multirow}
\usepackage{lipsum}

\begin{document}

\title{Electrostatic gate-controlled quantum interference in a high-mobility two-dimensional electron gas at the (La$_{0.3}$Sr$_{0.7}$)(Al$_{0.65}$Ta$_{0.35}$)O$_3$/SrTiO$_3$ interface}
\author{Km Rubi}
\email[email:] {rubi@lanl.gov}
\affiliation{National High Magnetic Field Laboratory, Los Alamos National Laboratory, Los Alamos, NM 87545, USA}

\author{Kun Han}
%\email{Present address: }
\affiliation{Department of Physics, National University of Singapore, 117551 Singapore}

\author{Huang Zhen}
%\email{Present address: }
\affiliation{Department of Physics, National University of Singapore, 117551 Singapore}

\author{Michel Goiran}
\affiliation{Laboratoire National des Champs Magnétiques Intenses (LNCMI-EMFL), Université de Toulouse, CNRS, INSA, UPS, 143 Avenue de Rangueil, 31400 Toulouse, France}

\author{Duncan K. Maude}
\affiliation{Laboratoire National des Champs Magnétiques Intenses (LNCMI-EMFL), Université de Toulouse, CNRS, INSA, UPS, 143 Avenue de Rangueil, 31400 Toulouse, France}

\author{Walter Escoffier}
\affiliation{Laboratoire National des Champs Magnétiques Intenses (LNCMI-EMFL), Université de Toulouse, CNRS, INSA, UPS, 143 Avenue de Rangueil, 31400 Toulouse, France}

\author{A. Ariando}
%\email{ariando@nus.edu.sg}
\affiliation{Department of Physics, National University of Singapore, 117551 Singapore}

\date{\today}

\begin{abstract}
We report quantum oscillations in magnetoresistance that are periodic in magnetic field ($B$), observed at the interface between (La$_{0.3}$Sr$_{0.7}$)(Al$_{0.65}$Ta$_{0.35}$)O$_3$ and SrTiO$_3$. Unlike Shubnikov–de Haas oscillations, which appear at magnetic fields $ > 7$ T and diminish quickly as the temperature rises, these $B$-periodic oscillations emerge at low fields and persist up to 10 K. Their amplitude decays exponentially with both temperature and field, specifying dephasing of quantum interference. Increasing the carrier density through electrostatic gating results in a systematic reduction in both the amplitude and frequency of the oscillations, with complete suppression beyond a certain gate voltage. We attribute these oscillations to the Altshuler–Aronov–Spivak effect, likely arising from naturally formed closed-loop paths due to the interconnected quasi-one-dimensional conduction channels along SrTiO$_3$ domain walls. The relatively long phase coherence length ($\sim$ 1.8 $\mu$m at 0.1 K), estimated from the oscillation amplitude, highlights the potential of complex oxide interfaces as a promising platform for exploring quantum interference effects and advancing device concepts in quantum technologies, such as mesoscopic interferometers and quantum sensors. 

 \end{abstract}
\maketitle

%\section{Introduction}

Quantum interference -- whether observed as the Aharonov-Bohm (AB) effect \cite{aharonov1959, webb1985}, the Altshuler–Aronov–Spivak (AAS) effect \cite{al1981, sharvin1981}, or quantum conductance fluctuations \cite{lee1985} -- offers valuable insights into electron dynamics within an electromagnetic environment. These phenomena allow direct measurement of the phase coherence length, which is critically important for developing quantum technologies such as quantum sensing, quantum computing, and quantum communication. The AB and AAS effects are manifested as conductance or resistance oscillations that are periodic in the magnetic flux, $\phi = BA$, where $B$ is magnetic field and $A$ is area enclosed by electron trajectory typically within ring or cylindrical geometries. These oscillations differ from Shubnikov–de Haas (SdH) oscillations, which are periodic in the inverse magnetic field ($1/B$) and arise from the Fermi energy crossing magnetically quantised electronic states. Notably, AAS oscillations persist even in the presence of disorder and survive ensemble averaging, making them robust in diffusive transport regimes. In contrast, AB oscillations are a hallmark of ballistic transport and tend to vanish when averaged over an ensemble of rings. Both AB and AAS oscillations have been observed in ring structures fabricated from III–V semiconductor heterostructure \cite{hansen2001, gul2014}, graphene \cite{russo2008, dauber2017}, and other two-dimensional (2D) materials.

The 2D electron gas (2DEG) at complex oxide interfaces offers a novel platform for exploring quantum interference in previously uncharted regimes. Unlike 2D electron systems in conventional III–V semiconductors, complex oxide interfaces such as LaAlO$_3$/SrTiO$_3$ (LAO/STO) host a strongly correlated 2DEG, leading to emergent phenomena like superconductivity \cite{caviglia2008, li2011, bert2011direct} and magnetism \cite{li2011, bert2011direct, brinkman2007, kalisky2012}. S. Goswami et al. fabricated superconducting quantum interference device at the superconducting LAO/STO interface and measured $B$-periodic oscillations in voltage by applying an excitation current higher than its critical value ~\cite{goswami2016}. Interestingly, AB-like oscillations were reported in a Hall bar device (channel width $w$ = 250 nm, length $l$ = 1.5 $\mu$m), but not in a nanoscale ring, both fabricated at the conducting LAO/STO interface  \cite{irvin2019}. 
%Further, the quasi-$B$-periodic oscillations in magnetoresistance observed at Gd/STO interface were attributed to classical Sondheimer oscillations \cite{mallik2021}. 

% using atomic force microscope lithography

To have a better understanding of the mechanism of $B$-periodic oscillations in complex oxide interfaces and their tunabilty through external knobs such as electrostatic gating, we studied quantum transport in an ultra-high mobility (35,000 cm$^2$V$^{-1}$s$^{-1}$) 2DEG at the interface between (La$_{0.3}$Sr$_{0.7}$)(Al$_{0.65}$Ta$_{0.35}$)O$_3$ (LSAT) and STO. Due to its smaller lattice mismatch ($\sim$ 1 $\%$) compared to LAO/STO ($\sim$ 3 $\%$), the LSAT/STO interface supports extremely high mobility 2DEG, reaching up to 50,000 cm$^2$V$^{-1}$s$^{-1}$ \cite{huang2016, han2017}. Furthermore, the LSAT/STO interface hosts long-range ordered line dislocations that form a 2D square moiré pattern with a moiré lattice constant of $\sim$ 40 nm~\cite{burian2021} and/or interconnected polar domain walls at low temperatures~\cite{Han2025}. These naturally formed defect lines may serve as 1D conduction channels for electron transport, providing a promising platform for studying quantum interference phenomena. 
%We observed both $B$-periodic quantum oscillations and conventional Shubnikov–de Haas (SdH) oscillations in Hall-bar patterned LSAT/STO devices. Both types of oscillations are tunable via electrostatic gating, though the $B$-periodic oscillations disappear beyond a certain positive gate voltage. We attribute the $B$-periodic oscillations to AAS interference, arising from naturally formed closed loops—either due to spatial inhomogeneities or a network of line dislocations at the interface. 

%The gate-tunability of these oscillations, along with the estimated coherence length of $\sim$ 1.3 $\mu$m, highlights a promising platform for investigating quantum interference phenomena in complex oxides and their potential applications in quantum technologies such as quantum interferometer. 

\begin{figure}[!htp]
\includegraphics[width=8.2cm]{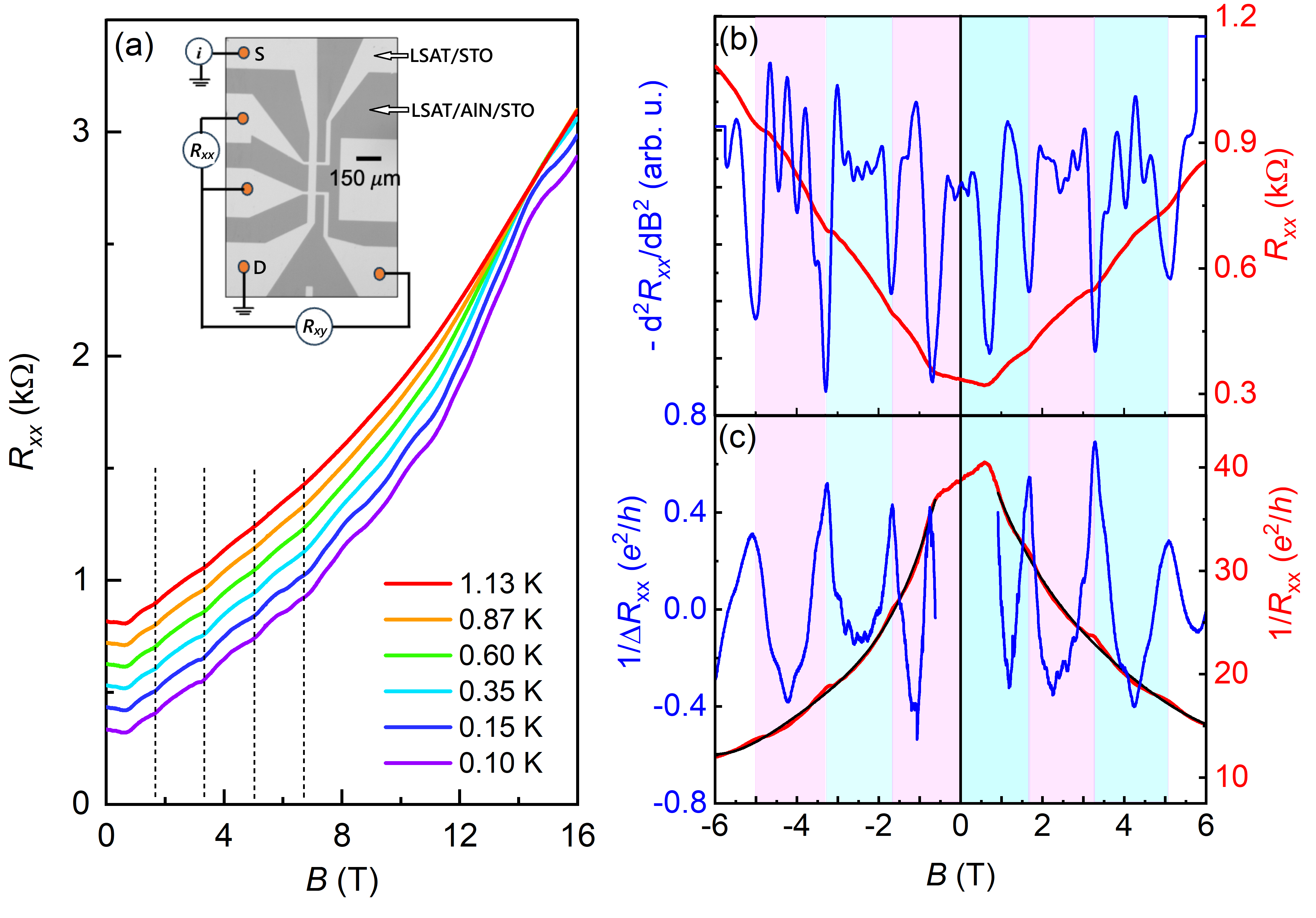} 
\caption{\label{F1} $B$-periodic oscillations in magnetoresistance suggesting quantum interference. (a) Magnetic field dependence of longitudinal resistance $R_{xx}(B)$ at a few selected temperatures and at a back-gate voltage $V_g$ = 3 V. The curves are shifted vertically with spacing of 0.1 k$\Omega$ for better visuality of oscillations. The $B$-periodic oscillations in $R_{xx}(B)$ in low fields (0 - 7 T), depicted by vertical dashed lines, survive above 1 K. Inset is a optical microscope image of the Hall-bar device with schematic of transport measurements scheme. (b) A second order derivative of $R_{xx}(B)$ (blue solid line and left y axis) reveals $B$-periodic oscillation along with higher harmonics. $R_{xx}(B)$ is displayed with red solid line and on right y axis. (c) Conductance (Inverse of $R_{xx}(B)$) in the unit of conductance quantum $e^2/h$ (right y axis) and the oscillating conductance after subtracting the background of $1/R_{xx}$ (left y axis). The black solid lines are the $2^{nd}$ order polynomial fit to the $1/R_{xx}$ data to subtract the monotonic background.}
\end{figure}

Here, we investigate magnetoresistance of LSAT/STO devices at temperatures down to 100 mK and in magnetic fields up to 16 T.  Our LSAT/STO devices consist of 10 unit cells of LSAT film grown on TiO$_2$-terminated STO (001) single crystals using pulse laser deposition (PLD) technique. The epitaxial growth of LSAT was precisely controlled by in-situ reflection high energy electron diffraction. During growth, the oxygen partial pressure (P${_{O_2}}$) and temperature were maintained at 1$\times10^{-4}$ torr and 950$^\circ$C, respectively. After growth, the samples were annealed in the oxygen atmosphere of pressure 25 mTorr and at the temperature of 600$^\circ$C for filling the oxygen vacancies. As displayed in Fig.~\ref{F1}a inset, the six-terminal Hall bar devices of width 50 $\mu$m and length 160 $\mu$m between the longitudinal voltage leads were fabricated by conventional photolithography technique using AlN films as hard mask. 
For transport measurements, the devices were electrically contacted with aluminum wires using wire bonder and then cooled from room temperature down to 100 mK in a dilution fridge. Simultaneous measurements of longitudinal resistance $R_{xx}$ and Hall resistance $R_{yx}$ were performed using an excitation current of 100 nA in static magnetic fields and 1 $\mu$A in pulsed magnetic fields. We measured two devices, device-1 and device-2. The measurements on the as-grown device-1 were not feasible due to extreme inhomogeneity leading to high resistance. To enable resistance measurements and tune carrier density on this device, we employed electrostatic gating through the STO substrate of thickness 0.5 mm. For back-gate operation, the rear surface of the substrate was coated with silver paint, and a gold wire was attached at the center of this surface to serve as the gate electrode. We performed measurements at different tilt angles with respect to $B$ orientation using in-situ rotator probe. Measurements at temperatures above 1.2 K were performed in a helium-4 cryostat.
% 

%\section{Experimental methods}

\begin{figure*}[!htp]
\includegraphics[width=18cm]{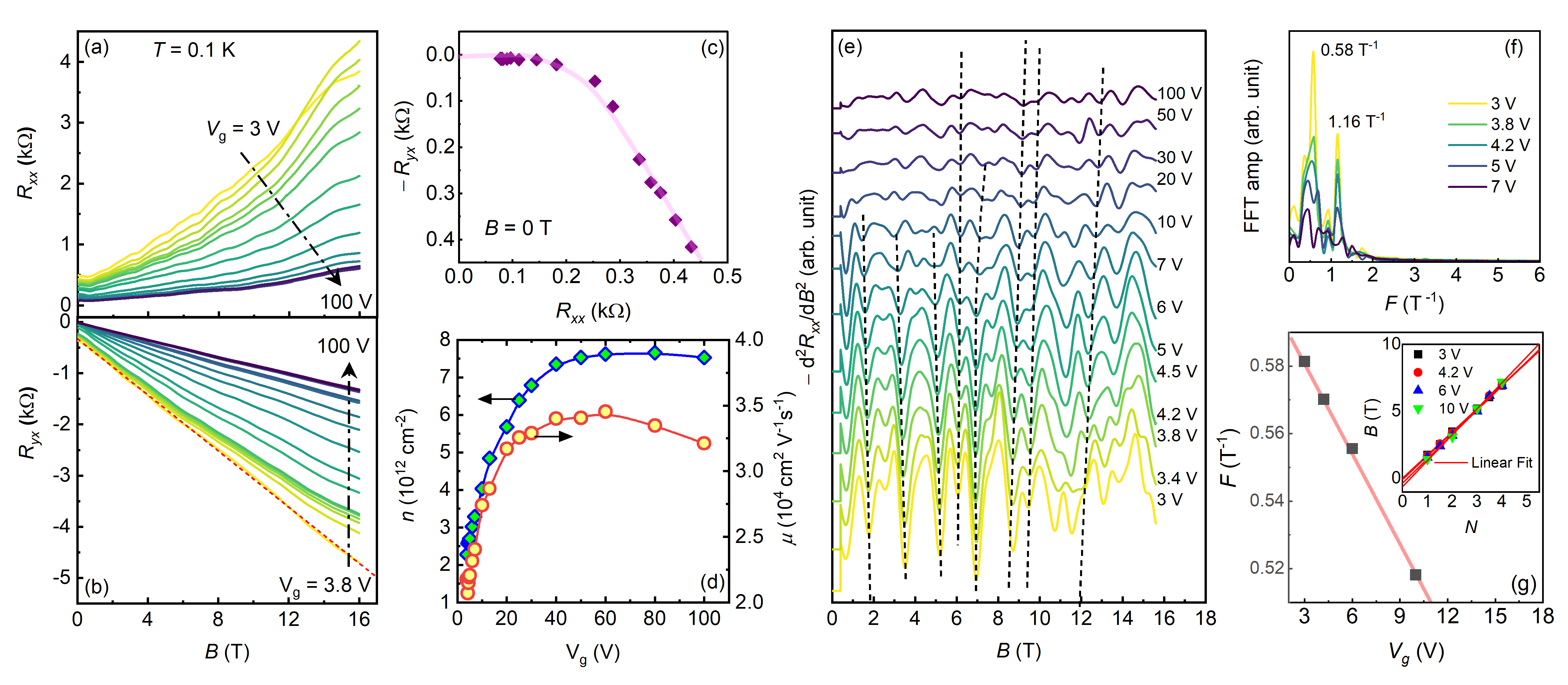} 
\caption{\label{Gating} Gating effect on electrical properties and quantum oscillations. (a) $R_{xx}(B)$ and (b) $R_{yx}(B)$ at different values of $V_g$. Dashed line in (b) is the linear fit to $R_{yx}(B)$ at $V_g$ = 3.8 V. (c) Zero-field offset of $R_{yx}$ as a function $R_{xx}$. Non-linear dependence of $R_{yx}$ on $R_{xx}$ in zero field indicates the existence of charge inhomogeneity at the interface. (d) Carrier density, $n$ (left y axis), and mobility, $\mu$ (right y axis) as a function of $V_g$. (e) An evolution of oscillations with varying $V_g$. $B$-periodic oscillations completely suppress for $V_g >$ 20 V. (f) FFT of $B$-periodic oscillations reveals two peaks, the main peak at 0.58 T$^{-1}$ and the harmonic peak at 1.16 T$^{-1}$. (g) Variation in the oscillations frequency with $V_g$. For the best estimation of frequency, we assign the minima to integer and linear fit the plot of minima positions vs their number. }
\end{figure*}

\begin{figure}[!htp]
\includegraphics[width=8.5cm]{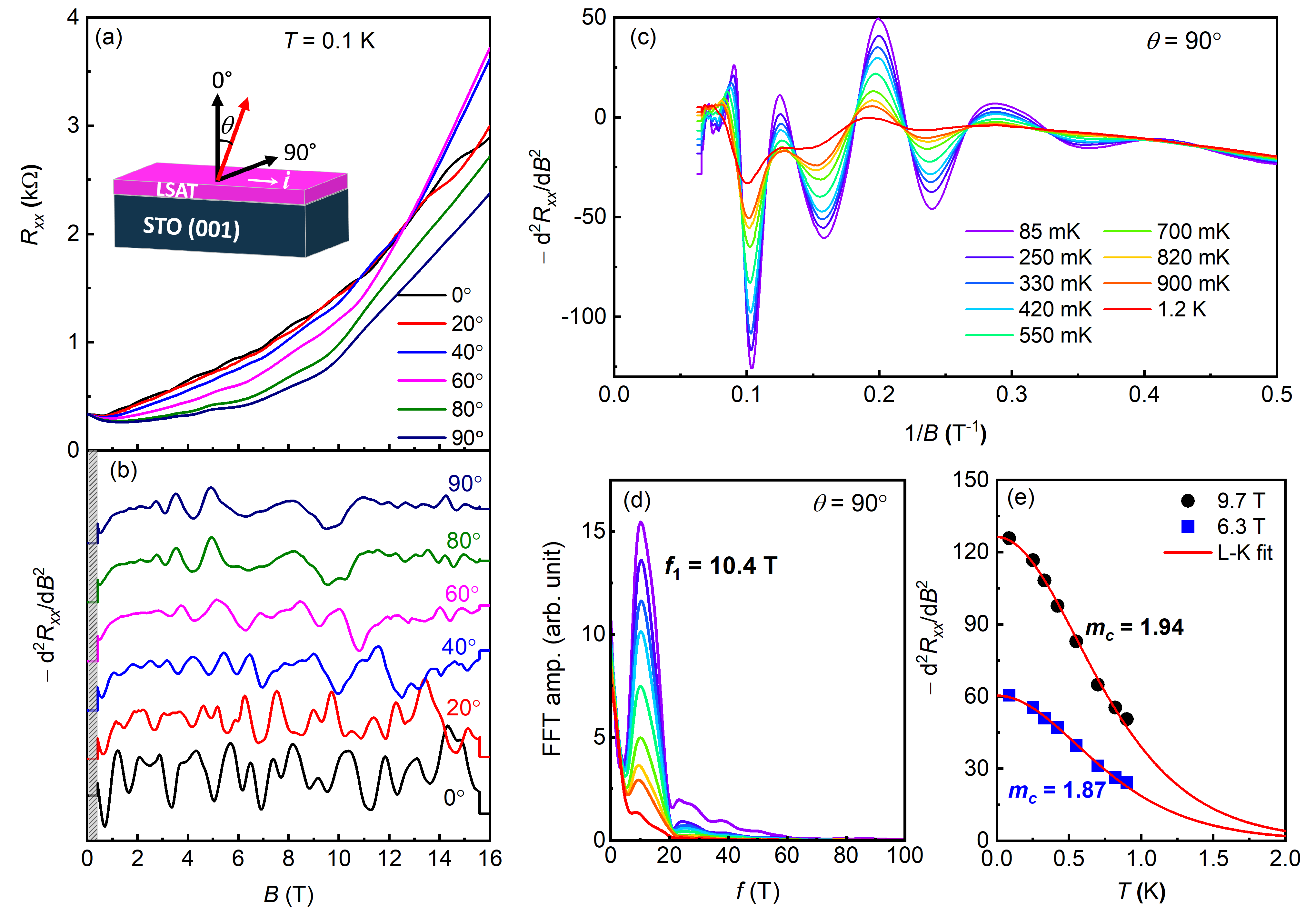} 
\caption{\label{Angle} Impact of $B$ orientation on quantum oscillations (a) $R_{xx}$ and (b) $-$d$^2R_{xx}$/d$B^2$ at different tilt angles measured at 0.1 K. The inset of (a) is a schematic of the magnetic field orientation, where $\theta$ = 0$^\circ$ and 90$^\circ$  are the angles when the field is perpendicular and parallel to the (001) plane, respectively. (c) Inverse field dependence of quantum oscillations measured at  $\theta$ = 90$^\circ$. (d) FFT spectra of oscillations at different temperatures. (e) Thermal damping of oscillations amplitude fitted with Lifshitz-Kosevich formula for cyclotron mass estimation.}
\end{figure}

\begin{figure}[!htp]
\includegraphics[width=8.2cm]{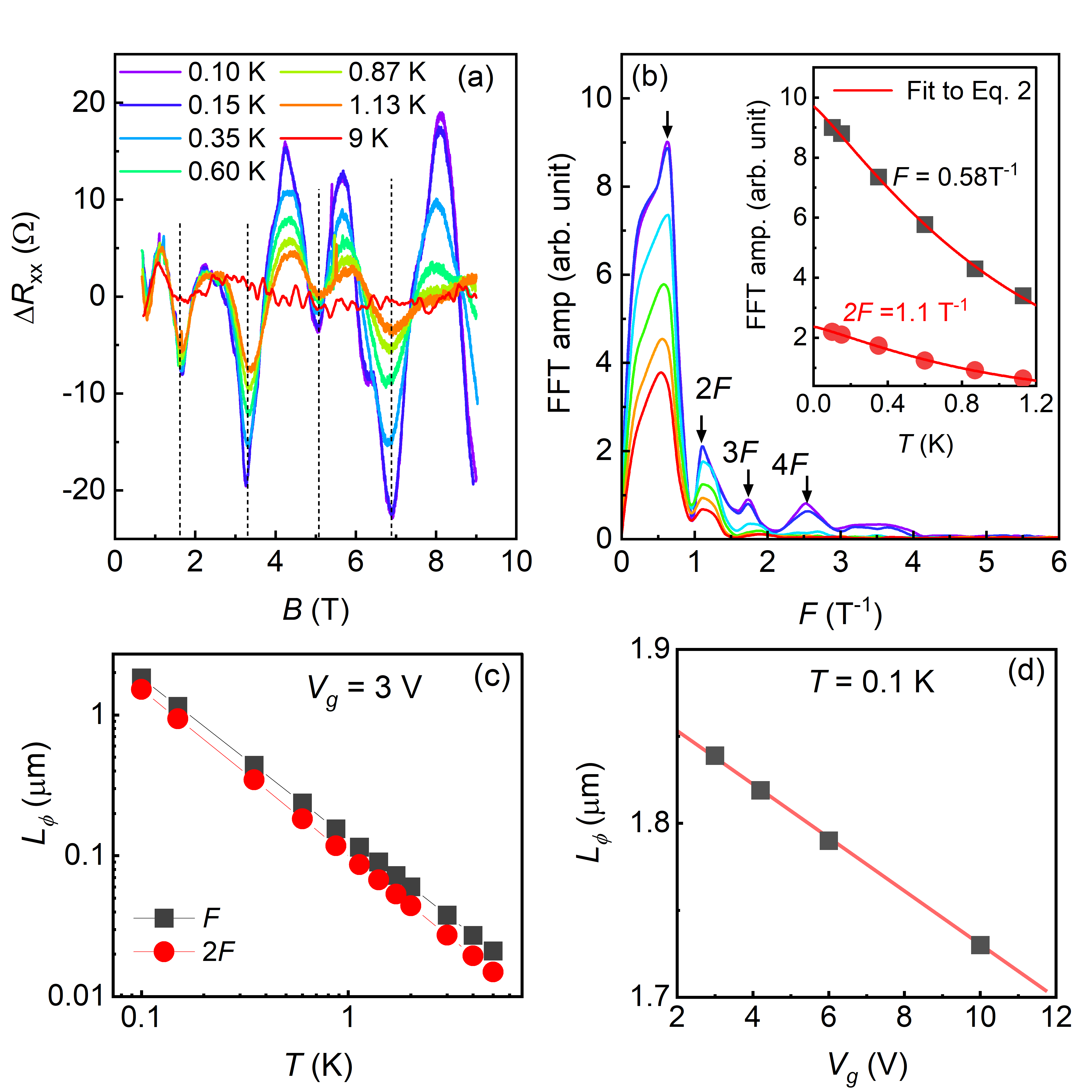} 
\caption{\label{F4} Phase coherence length $L_\phi$ estimation from the temperature dependence of $B$-periodic oscillations. (a) Oscillating component of  resistance at different temperatures. (b) Main panel: FFT of $B$-periodic oscillations producing 3 additional harmonic peaks. Inset: Temperature dependence of FFT amplitude with linear fit. (c) Temperature and (d) gate dependence of estimated $L_\phi$. }
\end{figure}

Figure~\ref{F1}a shows the longitudinal resistance, $R_{xx}(B)$, measured at various fixed temperatures and at a gate voltage of $V_g = $ 3 V for device-1. The positive magnetoresistance (MR) exhibits an oscillatory behaviour across the entire magnetic field range (0-16 T). Most interestingly, the oscillations in low fields (0-7 T) are periodic in $B$, as depicted by the vertical dashed lines, and persist up to temperatures of at least 9 K (see Fig.~\ref{F4} and Fig.~\ref{F5}). However, the high-field oscillation amplitude damps with increasing temperature and vanishes entirely above 1.2 K (see Fig.~\ref{SdH}).  While the high-field oscillations corresponds to the SdH effect, which is reported widely in STO-based 2DEGs \cite{ben2010, fete2014, mccollam2014, trier2016, rubi2020aperiodic, rubi2024}, the $B$-periodic oscillations observed at an oxide interface without a mesoscopic ring and ballistic transport are particularly intriguing. The second-order derivative of $R_{xx}(B)$ ($-$d$^2R_{xx}$/d$B^2$) in both field directions (up and down) reveals the existence of $B$-periodic oscillations of period $\sim$ 1.7 T, along with higher harmonics (Fig.~\ref{F1}b). Further, as displayed in Fig. 1c, the amplitude of these oscillations is $\sim$ 40 $\%$ of the quantum conductance ($e^2/h$) and decays as the magnetic field increases (Fig.~\ref{F5}).  

To understand the origin of $B$-periodic oscillations and investigate their tunability through electrostatic gating, we measured the same device at different values of gate voltage ($V_g$) at 0.1 K. Fig.~\ref{Gating}a and \ref{Gating}b show $R_{xx}(B)$ and $R_{yx}(B)$, respectively, for different $V_g$ values. Measurements of $R_{xx}(B)$ for $V_g <$ 3 V and $R_{yx}(B)$ for $V_g <$ 3.8 V were not possible because this as-grown sample was highly resistive due to extreme charge inhomogeneity. The inhomogeneity is evident from the $R_{yx}$ offset at zero magnetic field, which decreases progressively with increasing $V_g$ and approaches zero for $V_g \geq 20$ V. To identify if a misalignment in Hall voltage contacts is not a cause of $R_{yx}$ offset, we plot zero-field $R_{yx}$ as a function of $R_{xx}$ in Fig.~\ref{Gating}c. A non-linear dependence of $R_{yx}$ on $R_{xx}$ confirms that the $R_{yx}$ offset is due to the inhomogeneous distribution of carriers at the interface, not the misalignment of contacts.
 
From Fig.~\ref{Gating}b, the slope of $R_{yx}(B)$ decreases progressively with increasing $V_g$, indicating an increase in the electron population within the quantum well formed at the interface. This trend is further corroborated by a reduction in both the zero-field $R_{xx}$ and the positive MR (Fig.~\ref{Gating}a). We estimate the carrier density, $n$, through a linear fit to $R_{yx}(B)$ as shown by a dashed red line for $V_g$ = 3.8 V in Fig.~\ref{Gating}b, and plot $n$ and mobility  $\mu$ as a function of $V_g$ in Fig.~\ref{Gating}d. At $V_g$ = 3.8 V, $n$ is $2.3 \times 10^{12}$ cm$^{-2}$, which increases with $V_g$ up to 50 V and saturates at a maximum value of $\sim$ $7.5 \times 10^{12}$ cm$^{-2}$ beyond 50 V. This behaviour of $n(V_g)$ closely resembles previously reported trends at LAO/STO interfaces \cite{joshua2012, rubi2020aperiodic}. The saturation of $n$ with increasing $V_g$ can be attributed to the bending of the STO conduction band, which allows electrons to escape the quantum well and become trapped by defects \cite{biscaras2014limit, rubi2020aperiodic}. The mobility $\mu(V_g)$ follows the same trend as $n(V_g)$, except that the $\mu$ starts decreasing with $V_g$ above 60 V. For $V_g$ = 3.8 V, $\mu$ is $\sim$ 21,700 cm$^2$V$^{-1}$s$^{-1}$ and reaches a maximum value of $\sim$ 35,000 cm$^2$V$^{-1}$s$^{-1}$ at $V_g$ = 60 V. A linear proportionality of  $\mu$ with $n$ indicates the high quality of the electron gas, however, the decrease in $\mu$ above 60 V indicates a slight growth in ionised defects, which enhances the scattering rate while the density remains almost unchanged. 

As shown in Fig.~\ref{Gating}e and as a colour contour plot in Fig.~\ref{F6}, the oscillations exhibit a systematic shift in position as $V_g$ increases. Specifically, as indicated by the vertical dashed lines, the SdH oscillations at high fields ($B >$ 7 T) shift toward higher $B$, consistent with the chemical potential moving to higher energies with increasing positive $V_g$. In contrast, $B$-periodic oscillations, also marked by dashed lines, shifts toward lower $B$. Notably, the $B$-periodic oscillations amplitude diminishes with increasing $V_g$ and the oscillations nearly vanish at $V_g$ = 20 V, where the density exceeds $\sim 5 \times 10^{12}$ cm$^{-2}$ and zero-field $R_{yx}$ appraches zero. It is worth noting that the samples reported in Ref. \cite{han2017}, with densities on the order of $10^{13}$ cm$^{-2}$, did not exhibit $B$-periodic oscillations, consistent with the emergence of $B$-periodic oscillations only at lower carrier densities.

From the fast Fourier transform (FFT) analysis of $B$-periodic oscillations (Fig.~\ref{Gating}f), we extract a primary frequency of 0.58 T$^{-1}$ (corresponding to a period $B_0$ = 1.7 T) and a second harmonic at 1.1 T$^{-1}$. As displayed in Fig.~\ref{Gating}g, the frequency of the $B$-periodic oscillations decreases linearly with increasing $V_g$, a trend confirmed by both FFT analysis and a plot of the oscillation minima positions versus their index (inset of Fig.~\ref{Gating}g). 

As we know, our LSAT/STO device is not fabricated in a ring geometry; instead, it is a Hall bar with a length of 160~$\mu$m and a width of 50~$\mu$m. For a micro-scale spatially inhomogeneous sample, the transport mobility extracted using the Drude formula can be a rather crude estimate. To obtain a more reliable measure of electron scattering, we therefore estimate the quantum mobility by analyzing the SdH oscillations using the Lifshitz–Kosevich formalism~\cite{shoenberg2009magnetic}. The transport mobility is insensitive to small angle scattering, whereas quantum mobility is sensitive to all scattering events and is typically one order of magnitude smaller than transport mobility in 2D systems in which the carriers and ionized impurities are spatially separated. From the field-dependent of the SdH oscillations amplitude (see Fig.~\ref{SdH}b), we extract a quantum mobility of $\mu_q$ $\sim$ 2530 cm$^2$V$^{-1}$s$^{-1}$, one order of magnitude smaller than the transport mobility. Now, we calculate the Fermi wavelength as $\lambda_F = \sqrt{(2\pi/n)}$ = 16 nm, and the corresponding mean free path as $l = h\mu_q/e\lambda_F$ $\approx$ 65 nm (at $V_g$ = 3.0 V).
Comparing these values with the estimated circumference of the effective closed loop ($L$)  $\sim$123 nm (as $L = \sqrt{\frac{2\pi h}{eB_0}}$ and $B_0$ = 1.7 T), it is evident that the transport is not ballistic but diffusive. Therefore, the observed $B$-periodic quantum oscillations cannot be attributed solely to the AB effect.
Instead, these oscillations are likely a result of AAS interference, which typically arises from coherent backscattering along time-reversed paths in disordered systems or in networks of multiply connected loops \cite{umbach1986, naud2001}. Moreover, the presence of resistance maxima at $B$ = 0 across all temperatures (Fig.~\ref{F1}a) and exponential decay with magnetic field (Fig.~\ref{F5}b) further support the interpretation of AAS oscillations.

To verify the influence of magnetic flux on the $B$-periodic oscillations, we measured the same sample by rotating it with respect to magnetic field orientation at $T$ = 0.1 K, as illustrated in the inset of Fig.~\ref{Angle}a. In the low-field regime ($B$ < 12 T), the MR decreases with increasing tilt angle from 0$^\circ$ (out-of-plane field orientation) to 90$^\circ$ (in-plane field orientation); however, the MR shows a non-monotonic change in fields higher than 12 T.  More interesting is the evolution of oscillations with tilt angle as displayed in Fig.~\ref{Angle}b. The amplitude of all oscillations decays with increasing angle. The period of $B$-periodic oscillations reduces by $\sim 24 \%$ with increasing angle from 0$^\circ$ to 20$^\circ$, and these oscillations almost vanishes at 40$^\circ$. For large angles (60$^\circ$ - 90$^\circ$), new oscillations, whose amplitude and position do not dramatically change with field orientation, are resolved. These oscillations systematically follow thermal damping and almost vanish for $T >$ 1.2 K (Fig.~\ref{Angle}c). FFT analysis (Fig.~\ref{Angle}d) of $1/B$ dependence of oscillations reveals a single frequency of 10.4 T. The cyclotron mass is estimated from the fitting of temperature dependence of oscillations amplitude with the Lifshitiz-Kosevich (L-K) formula (Fig.~\ref{Angle}e) given below \cite{shoenberg2009magnetic}.

\begin{equation}
R(T) = R_0 \frac {2{\pi}^2k_Bm_cT/\hbar eB}{{\rm{sinh}}(2{\pi}^2k_Bm_cT/\hbar eB)} 
\end{equation}

The extracted mass $m_{\rm c}$ $\approx$ 1.9 $m_e$ indicates that these single-frequency oscillations originate from the quantization of 3D Fermi surface of STO \cite{bhattacharya2016}. The existence of 3D SdH oscillations suggests that a portion of the electrons reside deeper into the STO, consistent with previous observations at similar interfaces~\cite{rubi2024, yan2022ionic}. However, the influence of 3D carriers on the AAS oscillations is not yet well understood. Based on our current understanding, we expect the transport contributions from 2D confined channels at the interface and from 3D bulk-like channels in STO to be largely decoupled. Both in our previous work ~\cite{rubi2024} and in the present study, we find that SdH oscillations associated with 3D carriers are significantly different than those originating from 2D carriers, indicating that when the magnetic field is applied perpendicular to the interface, electrical conduction and the resulting quantum interference effects are dominated by 2D carriers.

Next, to understand the origin of the observed AAS effect in our Hall bar-patterned LSAT/STO, we consider two possible scenarios for the existence of naturally formed closed-loop paths: (1) a 2D network of line dislocations forming a square moiré pattern, resulting from complete lattice relaxation at the LSAT/STO interface~\cite{burian2021}; and (2) interconnected domain walls originated below cubic-to-tetragonal phase transition of STO~\cite{Han2025, kalisky, honig, ma2016}. If the 2DEG is spatially inhomogeneous—as indicated by the gate-tunable zero-field offset observed in $R_{xy}$ (Fig.~\ref{Gating}b and c)—then conduction through interconnected quasi-1D channels can occur in either scenario. In the naturally created network of such closed-loop paths, phase-coherent electrons can traverse time-reversed paths around loops, leading to constructive interference. However, the formation of a 2D moiré pattern is an intrinsic property of the LSAT/STO interface and should lead to a consistent oscillation period across samples grown under similar conditions. In contrast, STO domain walls form randomly, and their configurations are not expected to be identical between samples, leading to variations in oscillation periodicity. To test this, we examined another LSAT/STO device (device-2) grown in same conditions as device-1 and observed oscillations with a period of approximately 3 T (Fig.~\ref{F7}), nearly twice that of the device-1. This significant difference indicates that the $B$-periodic oscillations in LSAT/STO originate from conduction through randomly arranged and sufficiently interconnected STO domain walls, driven by spatial inhomogeneity. Further evidence for domain walls formation is provided by the temperature dependence of the resistance. $R_{xx}(T)$ of device-1 measured while cooling from room temperature to 1.5 K (Fig.~\ref{RT}a), exhibits two distinct anomalies at $\sim$ 100 K and $\sim$ 40 K. The anomaly near 100 K is widely attributed to the cubic-to-tetragonal structural transition in STO \cite{Minhas2017, Kwak2022, Han2025}. And, the anomaly around 40 K, which shows pronounced hysteresis between cooling and warming cycles (Fig.~\ref{RT}b) and strong gate dependence \cite{Han2025}, is attributed to the formation of polar domain walls.

To estimate the phase coherence length ($L_\phi$), we further analyzed the temperature dependence of the $B$-periodic oscillations of device-1. After subtracting a monotonic background from the data in Fig.~\ref{F1}a, we plot the oscillatory component of the resistance at different temperatures in Fig.~\ref{F4}a. As expected, the oscillation amplitude—and correspondingly, the FFT amplitude (Fig.~\ref{F4}b)—decreases with increasing temperature, consistent with the suppression of phase coherence with temperature. In a closed loop of circumference $L$,  the oscillations amplitude decays exponentially with the ratio $L/L_\phi$ \cite{hansen2001}: 
\begin{equation}
A(T) = A_0 \exp \left( -N\frac{L}{L_\phi(T)}\right)
\end{equation}
where $L_\phi(T) \propto T^{-p}$ and $N$ (= 1, 2....) is the number of harmonics. 
%From Eq. (1), 
%\begin{equation}
%\ln \left(\frac{A(T)}{A_0}\right) = -N\frac{L}{L_\phi(T)}
%\end{equation}
The temperature dependence of the FFT amplitude for the first two harmonics (inset of Fig.~\ref{F4}b) fits well with Eq. 2 for $p$ = 1.14, indicating the suppression of quantum interference due to charge fluctuations \cite{seelig2001}. Using the fitting parameters and estimated circumference of the effective closed loop ($L$)  $\sim$123 nm, we calculate $L_\phi$ = 1.84 $\mu$m at 0.1 K for the first harmonic, which is twice as large as that for a LAO/STO device with channel lengths nearly 100 times shorter \cite{irvin2019}. We show the temperature dependence of $L_\phi$ from the first two harmonics in Fig.~\ref{F4}c. The extracted exponent $p \sim 1.1$ suggests that decoherence in LSAT/STO closely resembles that in ballistic mesoscopic systems, where $L_\phi(T) \propto T^{-1}$ and dephasing is primarily driven by electron-electron interactions \cite{hansen2001, dauber2017}. 

Assuming a similar temperature dependence of the oscillation amplitude at other $V_g$, and using the oscillation frequencies shown in Fig.~\ref{Gating}g, we estimate the $L_\phi$ for various $V_g$ values. $L_\phi$ decreases almost linearly with increasing $V_g$ (Fig.~\ref{F4}d), following the trend in oscillation frequency. It is worth noting that the amplitude of the $B$-periodic oscillations also diminishes with increasing $V_g$, and these oscillations vanish completely for $V_g > 20$ V. This behavior can be attributed to the increase in carrier density with $V_g$, which reduces the Fermi wavelength $\lambda_F$. A shorter $\lambda_F$ allows more wave modes to contribute with different phases, resulting in partial dephasing and a reduction in oscillation amplitude due to mode averaging. Similar suppression of oscillations amplitude with increasing $V_g$ has been reported in GaAs/AlGaAs heterostructures \cite{ford1989} and graphene \cite{dauber2017}. The simultaneous reduction in both oscillation frequency and amplitude with increasing $V_g$ can also be attributed to a decrease in spatial inhomogeneity. As carriers accumulate at the interface, the charge distribution becomes more uniform, reducing the size of naturally-formed closed-loop paths. Eventually, this evolution will lead to the disappearance of interconnected closed-loop paths at some point and completely vanish $B$-periodic oscillations.  

The characteristic mesoscopic length scale extracted from the observed oscillations is on the order of $0.1~\mu$m, raising the question of whether it corresponds to the STO domain size and whether AAS oscillations can persist in the presence of spatially varying domains. Although we do not have a direct measurement of ferroelastic domain sizes in LSAT/STO, studies on LAO/STO interfaces report domain widths ranging from $0.1~\mu$m - $10~\mu$m, depending on strain, thermal history, polarization, and carrier density. A direct one-to-one correspondence between the oscillation-derived loop size and the macroscopic domain spacing is therefore not expected. Importantly, AAS oscillations can survive spatial variations in domain size or disorder because they originate from constructive interference between time-reversed electron trajectories, whose phases remain identical under ensemble averaging. Consistent with this picture, $B$-periodic oscillations have been observed in multiwalled carbon nanotubes despite a wide distribution of tube diameters ranging from 1 to 20 nm \cite{Bachtold1999, Fujiwara1999, Kang2008}. Moreover, theoretical studies of AAS oscillations in disordered networks and granular conductors show that although the underlying loop-area distribution may be broad, the effective contribution is narrowed by phase-coherence constraints and magnetic-field averaging \cite{Sharvin1984}. Using the empirical relation $\Delta A/A=B_0/B_{max}$, where $B_{max}$ (= 9 T) is the field where the oscillations completely vanish and $B_0$ (= 1.7 T) is the period of oscillations for device-1, we estimate an effective relative loop-area dispersion of $\sim 0.2$, consistent with theoretical expectations for random network systems. Future studies such as direct imaging of domain walls at LSAT/STO interfaces with scanning probe techniques could further test the proposed connection between mesoscopic transport and naturally-formed domain structures and provide a better understanding of the microscopic origin of the robust $B$-periodic oscillations.

In conclusion, we have fabricated high-mobility Hall-bar-patterned LSAT/STO devices and performed magnetotransport measurements down to 0.1 K. At low magnetic fields (0–7 T), the devices exhibit a positive MR superimposed with pronounced quantum oscillations that are periodic in magnetic field $B$, indicating fully phase-coherent transport. The amplitude of these $B$-periodic oscillations decays exponentially with increasing temperature and magnetic field strength, consistent with dephasing of quantum interference. We attribute the $B$-periodic oscillations to AAS interference, most likely arising from naturally formed closed-loop paths due to the spatial inhomogeneity, which supports high-mobility conduction along the interconnected SrTiO$_3$ domain walls. The observed gate-tunability, combined with the relatively long phase coherence length ($\sim$ 1.8 $\mu$m at 0.1 K) underscores the potential of complex oxide interfaces as a platform for studying quantum interference phenomena and for developing quantum technologies such as mesoscopic interferometers and quantum sensors. 

Acknowledgements: K. R. acknowledge support from the National High Magnetic Field Laboratory, supported by the National Science Foundation through NSF/DMR-2128556 and the state of Florida and the US Department of Energy (DoE). This work was performed as part of the DoE BES project "Science of 100 Tesla". This study has been partially supported through the EUR grant NanoX no. ANR-17-EURE-0009 in the framework of the "Programme des Investissements dAvenir". This work at NUS is supported by the Ministry of Education (MOE), SIngapore, under its Tier-2 Academic Research Fund (AcRF) and by the MOE Tier-3 Grant (MOE-MOET32023-0003) 'Quantum Geometric Advantage'. 

\section*{Appendices}

\renewcommand{\thesubsection}{A\arabic{subsection}}

\setcounter{subsection}{0}

\renewcommand{\theequation}{A\arabic{equation}}

\setcounter{equation}{0}

\begin{figure}[!htp]
\includegraphics[width=8.5cm]{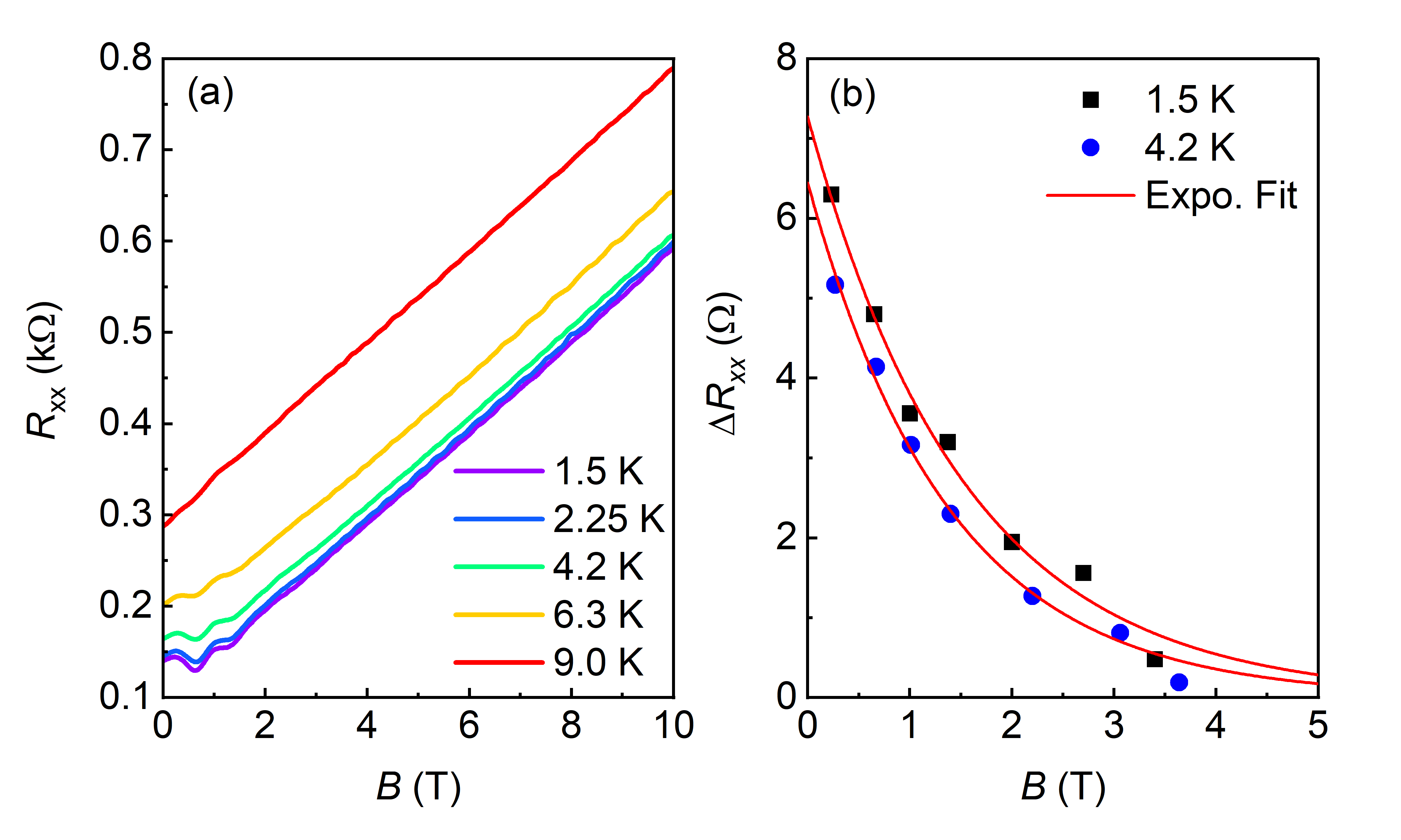} 
\caption{\label{F5} (a) Damping of $B$-periodic oscillations with magnetic field and temperature. No oscillations are observed above 9 K (b) Oscillations follow exponential decay with magnetic fields.}
\end{figure}

\emph {Appendix 1} To investigate the temperature range over which the $B$-periodic oscillations persist, we measured the device-1 in a helium-4 cryostat at temperatures ranging from 1.5 K to 9.0 K at $V_g = 3.0~$V. These measurements were carried out in pulsed magnetic fields of pulse length $\sim$ 300 ms. Fig.~\ref{F5}a display $R_{xx}(B)$ measured at different temperatures. The zero-field $R_{xx}$ from these measurements differs slightly, consistent with changes in electrical conduction commonly observed in STO-based interfaces after thermal cycling. Such changes are often attributed to oxygen vacancy redistribution \cite{Minhas2017, goble2017}, leading to slight variation in carrier density and mobility.  $B$-periodic oscillations remain visible at least up to 9 K (Fig.~\ref{F5}a); and the amplitude of these oscillations exhibits an exponential decay with increasing magnetic field strength (Fig.~\ref{F5}b). 

\begin{figure}[!htp]
\includegraphics[width=8.5cm]{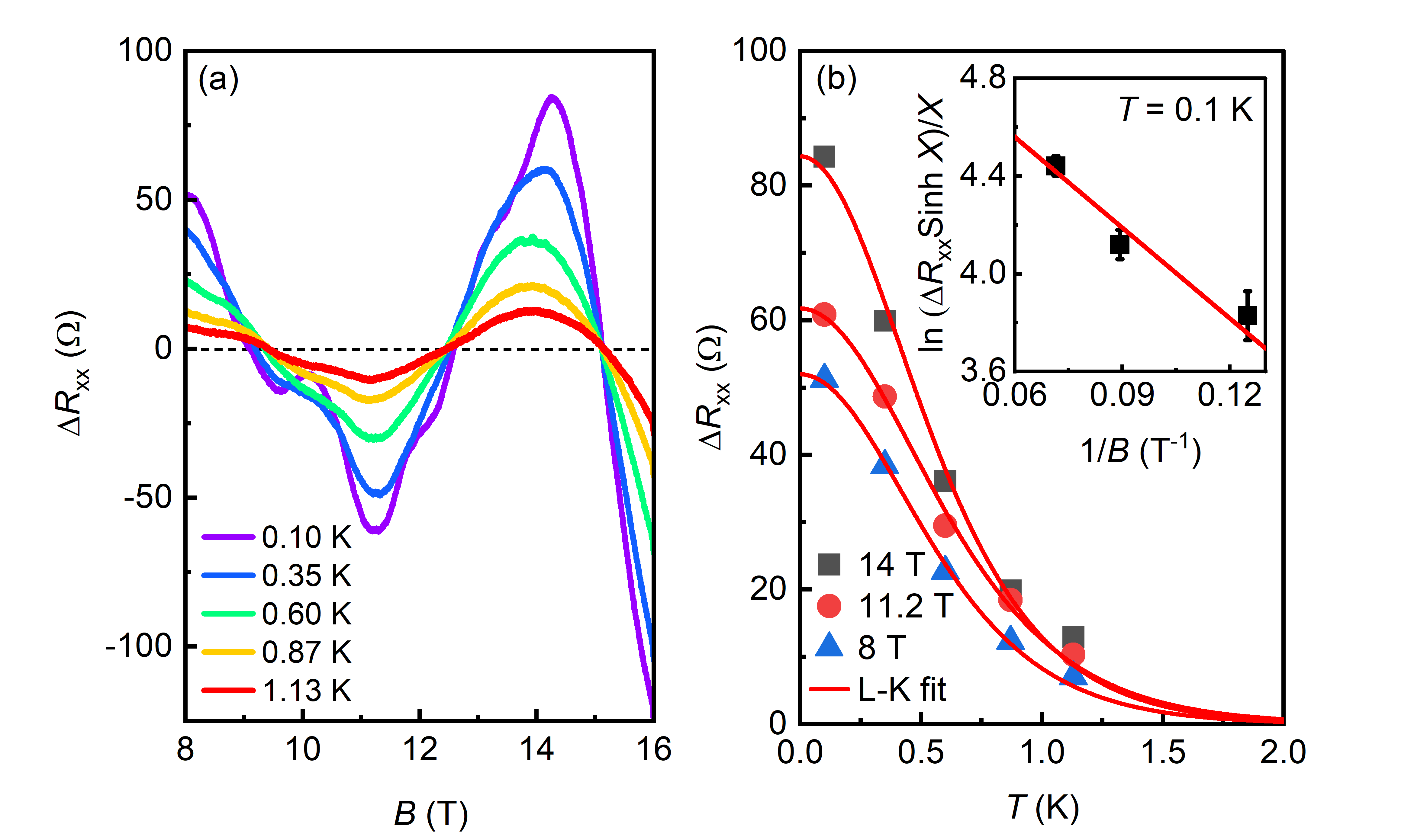} 
\caption{\label{SdH} (a) SdH oscillations after subtracting the monotonous background of $R_{xx}(B)$. (b) Main panel: Temperature dependence of SdH oscillations amplitude $\Delta R_{xx}$ at different magnetic fields fitted with L-K formula for cyclotron mass extraction. Inset: A linear fit to the inverse field dependence of normalized $\Delta R_{xx}$ for the estimation of quantum mobility.}
\end{figure}

\begin{figure}[!h]
\includegraphics[width=8.5cm]{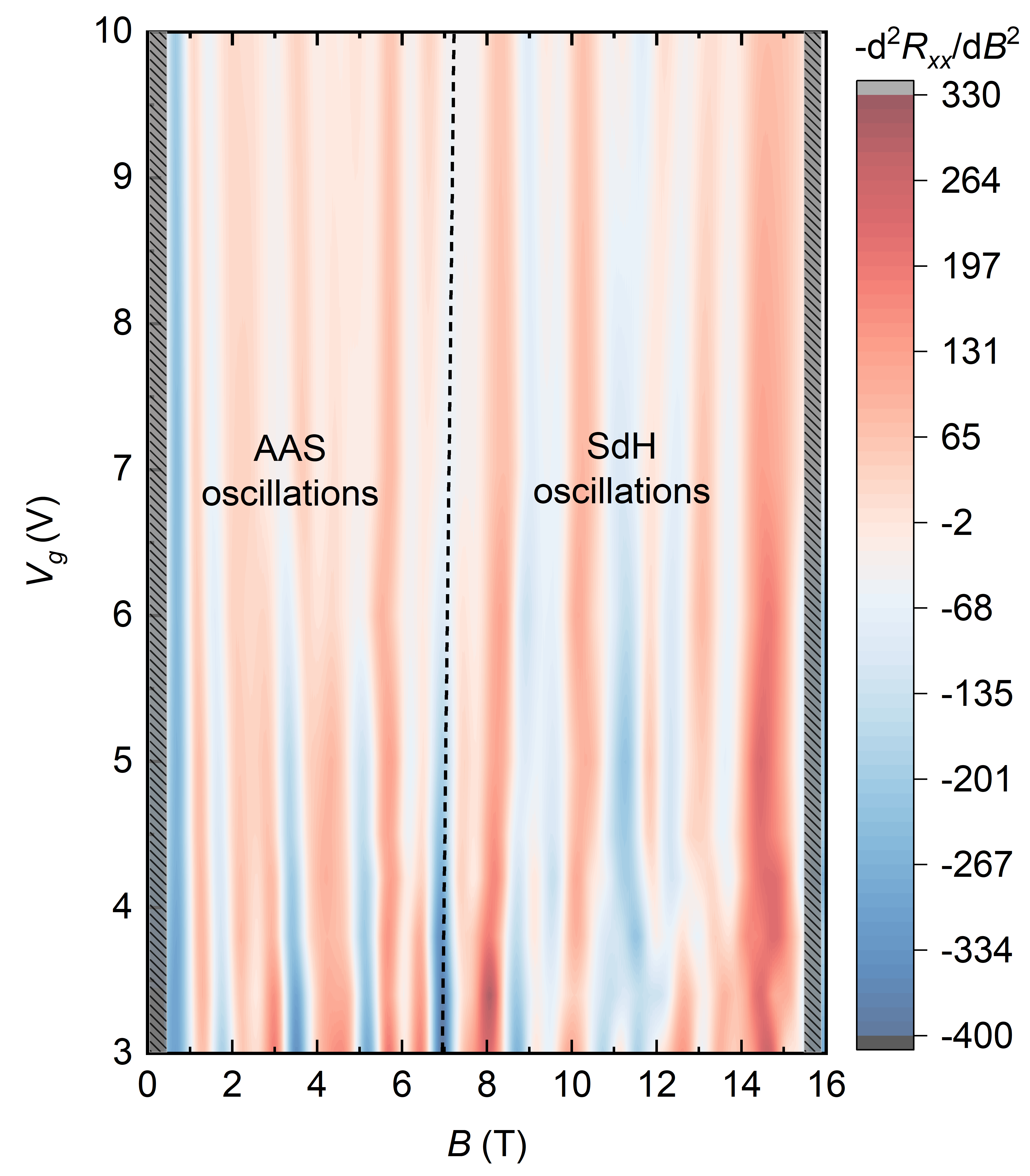} 
\caption{\label{F6} Colour contour plot of oscillations evolution with $V_g$ at $T$ = 0.1 K. Blue colour represents oscillations minima and red maxima. The vertical dashed line at $B \sim$ 7 T divide the region of $B$-periodic AAS oscillations and SdH oscillations. The decay of $B$-periodic oscillations and their shift to lower $B$ with increasing $V_g$ is clearly visible. However, we don't see any amplitude change with $V_g$ for SdH oscillations.}
\end{figure}

\emph {Appendix 2} To analyse the SdH oscillations observed for $\theta = 0^{\circ}$, we subtracted the monotonous background of $R_{xx}(B)$ data displayed in Fig.~\ref{F1}(a). At the lowest temperature 0.1 K, we observed large amplitude low-frequency oscillations superimposed with higher frequency oscillations (Fig.~\ref{SdH}a). As the temperature increases, the high-frequency oscillations start damping quickly and the low-frequency oscillations survive at least up to $\sim$ 1.1 K. By fitting the temperature dependence of oscillations amplitude with L-K formula as shown in the main panel of Fig.~\ref{SdH}b, we estimated the cyclotron mass of $\sim 2.1 m_e$. Further, analyzing the exponential growth of oscillations with $B$ (inset of Fig.~\ref{SdH}b), we extracted the quantum mobility of $\sim 2530$ cm$^2$V$^{-1}$s${-1}$, one order of magnitude smaller than the transport mobility estimated from Drude formula.

\emph {Appendix 3} To further illustrate the evolution of quantum oscillations with $V_g$, we replot the data from Fig.~\ref{Gating}e as a color contour map (Fig.~\ref{F6}). In this representation, the blue regions correspond to oscillation minima and the red regions to maxima. This format clearly reveals a shift of the $B$-periodic oscillations (AAS oscillations) toward lower magnetic fields and SdH oscillations toward higher magnetic fields with increasing $V_g$.

\begin{figure}[!htp]
\includegraphics[width=6.0cm]{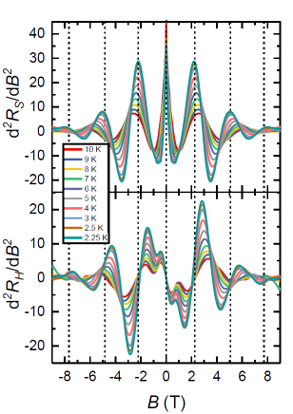} 
\caption{\label{F7} $B$-periodic oscillations in the longitudinal resistance (top) and Hall resistance (b) of device-2. The periodicity of oscillations is $\approx$ 3.0 T. These measurements were conducted without any electrostatic gating.} 
\end{figure}

\begin{figure}[!htp]
\includegraphics[width=8.5cm]{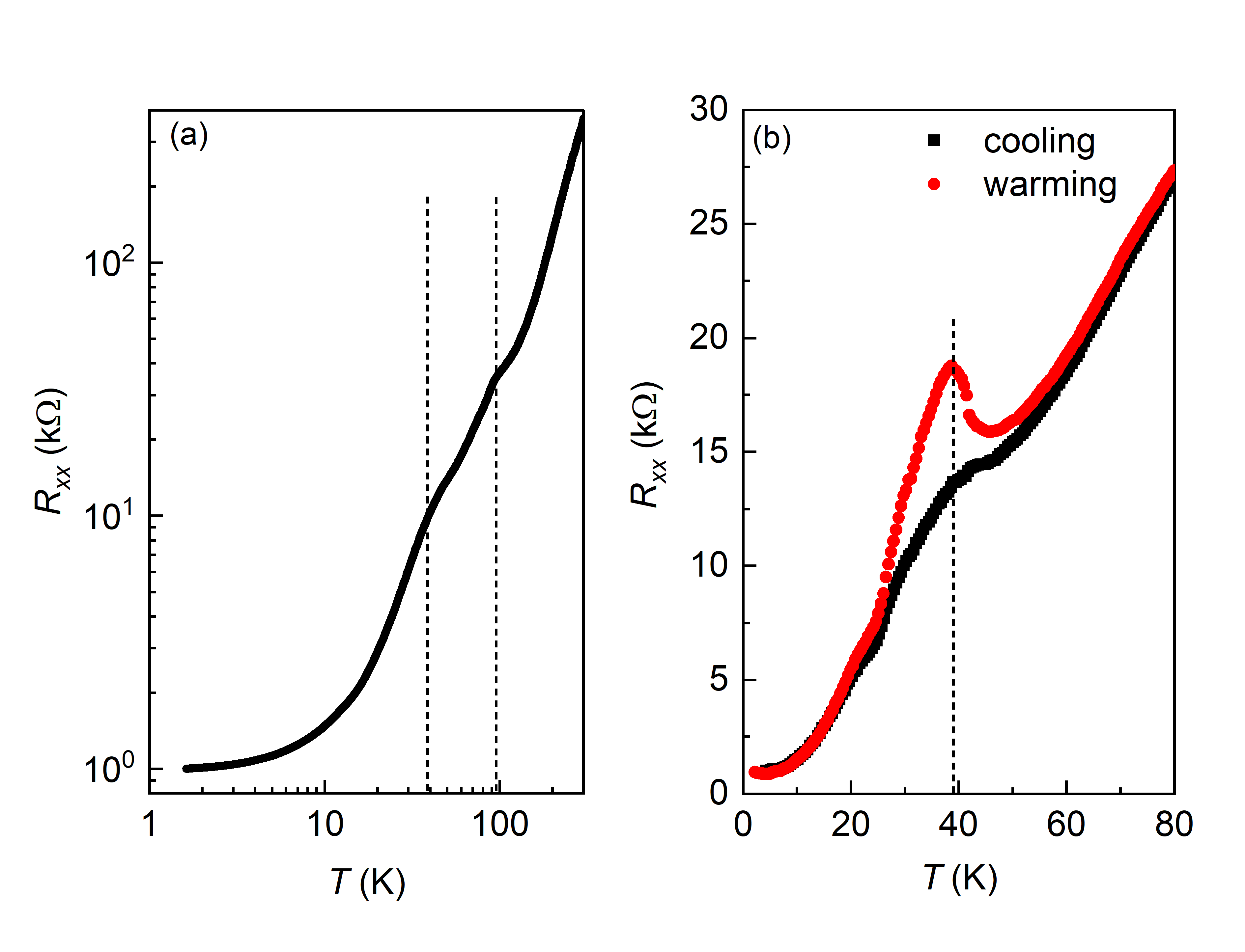} 
\caption{\label{RT} Temperature dependence of resistance $R(T)$ of device-1. (a) $R(T)$ in the temperature range of 300 - 1.5 K measured while cooling. (b) Comparison of $R(T)$ measured while cooling and warming in the low temperature regime. The dashed vertical lines depict the anomalies observed in $R(T)$.}
\end{figure}

\newpage

%\bibliographystyle{apsrev4-2-Title}
%\bibliography{Manuscript}

%

\end{document}